# The Efficiency of Quantum-Mechanical Carnot Engine using the Woods-Saxon model.

E. O. Oladimeji[†1], T. T. Ibrahim[1], A. N. Ikot[2], J. D. Koffa[1], H. O. Edogbanya[3], E. C. Umeh[1],

J. O. Audu[1], J. M. Uzer[1].

1. Theoretical Physics Group, Department of Physics, Federal University Lokoja, Lokoja, Nigeria.
2. Theoretical Physics Group, Department of Physics, University of Port Harcourt, Port Harcourt, Nigeria.
3. Department of Mathematics, Federal University Lokoja, Lokoja, Nigeria.

## ABSTRACT

The quantum engine cycle serves as an analogous representation of classical heat engines for microscopic systems and the quantum regime of thermal devices is composed of a single element. In this work, the Quantum-Mechanical properties of a non-linear quantum oscillator described by the Woods-Saxon [WS] model are examined. The Quantum-Mechanical analogue of the Carnot cycle was constructed using changes in both the width $L$ of the well and the quantum state of the potential well. The efficiency of the quantum engine, consisting of adiabatic and isothermal processes based on the Woods-Saxon [WS] potential is derived. The result is shown to be analogous to that of the classical engine and found to agree, within an appropriate limit, with existing results obtained from other potential models. This implies that the [WS] potential can be used as an alternative model in quantum engines.

---

[†] Corresponding author: E. O. Oladimeji. e-mail: nockjnr@gmail.com

# 1. INTRODUCTION

The study of the relation between heat engines and quantum systems as a working mechanism, which is often referred to as Quantum Heat Engines (QHEs) has always been of keen interest [1–12] since the concept was introduced in 1959 by *Scovil and Schultz-Dubois* [13]. Generally, heat engines are known to convert heat energy $Q_H$ to useful work $W$ and its performance quantifiers are the efficiency $\eta$ and the power output $P$ [14]. Practically, however, there is a delicate balance between these quantities such that their peaks are not attainable due to heat leaks and/or loss of energy through friction [15,16], which in turn devalues the efficiency $\eta$ of the engine such that $\eta \leq \eta_c$ [17], where the Carnot efficiency $\eta_c = 1 - T_C/T_H$ given that $T_C$ is the (cold)low temperature in the heat reservoir and $T_H$ the high temperature in the heat reservoir. However, these concerns become insignificant with the QHEs, since its dynamical equations of motion obey the laws of quantum mechanics [18].

Over fifty years of extensive theoretical studies of the miniaturization of heat engine from macroscale to nanoscale has shown that the limitations of classical heat engines become negligible as its quantum properties become dominant at the latter scale [19,20]. These studies have employed appropriate quantum equations of motion together with different working substances such as the spin systems [21,22], two-level or multilevel systems [23,24], particle in a box [FP] [24,25], cavity quantum electrodynamics systems [26,27], coupled two-level systems [28], Harmonic oscillators [HO] [18,29], Pöschl-Teller Oscillator[PTO] [30–32], Shortcut-to-Adiabacity [33–35] etc. The spin-offs of these earlier works are the fabrication of classical micro heat engines using optomechanical [36], micro-electromechanical [10,37], and colloidal systems [8,38], however, to date no quantum heat engine has been constructed.

Following the quantum mechanical model of the cyclic engine in which a single particle is confined in an infinite one-dimensional potential well of width $L$ as proposed by *Bender et al* [24], the temperature $T$ is replaced by the expectation value of the Hamiltonian $E$, that is, the ensemble average of the energies of the quantum particle and the walls of the containing potential play the role of the piston in a cylinder containing an ideal gas and the quantum mechanical equivalents of isothermal and adiabatic processes were constructed. Their concept was useful in other works where they formulated and applied the Pöschl-Teller [PT] potential in Carnot [30], Joule-Brayton and Otto cycles [32], rather than the [FP] model used by *Bender et al*.

In this paper, we introduce the Woods-Saxon [WS] model to construct an adiabatic and isothermal quantum analogous process and analyze the efficiency of an idealized reversible heat engine. This potential was introduced in 1954 by *R.D. Wood* and *D.S. Saxon* for studying nuclear structure and reaction properties [39], however its application in Quantum systems until recently has not been studied extensively. Its application in simulating confinement of Quantum Dots [QDs] was first proposed by *Costa et al* [40] where they proved the [WS] model to be a good alternative for studying confined quantum systems. *Xie* presented theoretical analysis of an exciton and studied two electrons confined in a quantum dot with the Woods-Saxon potential in [41] and [42] respectively. *Lu et al.* observed the effect of an intense, high-frequency laser field on the optical absorption coefficients and the refractive index change for shallow donor impurities in a [QD] using the [WS] potential [43]. *Aytekin et al.* proved that the [WS] potential gives good description of the properties of Nano-systems since it can be used to model the confinement in quantum systems with considerable success [44]. Hence, the [WS] model has been shown to be suitable for studying nonlinearities in quantum systems and would be of good interest to understand the performance of [QHE] whose working substance is constrained within this more realistic potential model.

The one-dimensional form of the potential is given by

$$V(x) = \frac{-V_0}{1 + e^{\left(\frac{x-L}{a}\right)}}$$

where $V_0$ is the depth, $L$ is taken to represent the confinement width and $a$ is the diffuseness of the interaction. The corresponding quantized energy spectrum takes the form [45]:

$$E_n(L) = -\left[\frac{V_0}{2} + \frac{(n+1)^2 \hbar^2}{8mL^2} + \frac{V_0^2 mL^2}{2\hbar^2(n+1)^2}\right] \qquad 1$$

where $m$ is the mass of the particle or the working substance.

## 2. THE CARNOT CYCLE

The Carnot cycle is an ideal model of a reversible heat engine with maximum possible efficiency. It is widely accepted as the most efficient cycle attainable by physical law. The cycle is composed of two thermodynamic reversible processes namely, isothermal and adiabatic processes. In a classic engine, the system remains in equilibrium throughout an isothermal process i.e., both the temperature $T$ and the internal energy of the gas remain constant, and work is done as the system remains in contact with the heat source while the gas expands. In the quantum mechanical representations, the initial state $\psi(x)$ is assumed to be a linear combination of eigenstates $\phi_n(x)$, the expectation value of the Hamiltonian remains constant while the size of the potential well changes with the wall's movement. Thus, the expansion coefficient $a_n$ changes such that $E(L)$ remains fixed as $L$ changes:

$$E(L) = \sum_{n=1}^{\infty} |a_n|^2 E_n \qquad 2$$

where $E_n$ is the energy spectrum and the coefficients $a_n$ satisfies the normalization condition $\sum_{n=1}^{\infty} |a_n|^2 = 1$. The pressure $P$ exerted on this wall as defined by Hellman and Feynman as [46,47]:

$$\hat{P}_n(L) = -\frac{\partial \hat{E}_n}{\partial L} \qquad 3$$

where $\hat{E}$ is defined as the energy operator such that the formal relation between pressure $\hat{P}$ and Hamiltonian is $\hat{P}(\hat{x}, \hat{p}, L) = -(\partial/\partial L)\hat{H}(\hat{x}, \hat{p}, L)$.

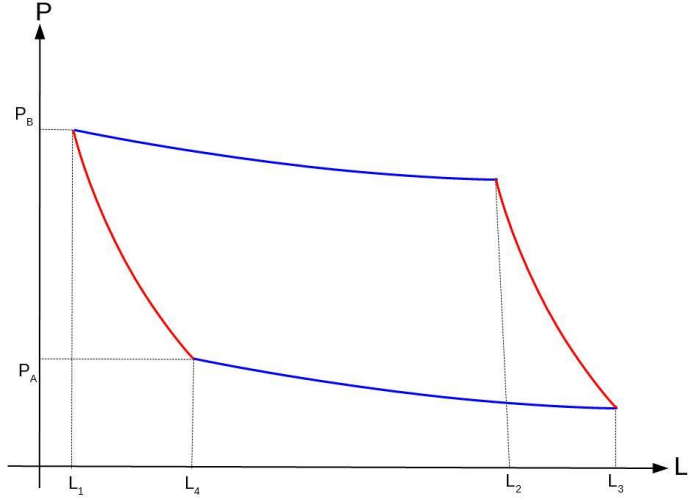

*Figure 1: The schematic representation of Carnot's cycle, drawn in (P, L) coordinates* [30].

a. **Process 1: Isothermal Expansion**

In the first process, the piston is set into an isothermal expansion as the system is excited from its initial state $n = 1$ to the second state $n = 2$ (i.e., from $L = L_1$ to $L = L_2$). This process keeps the expectation value of the Hamiltonian constant and as a result, the state of the system is a linear combination of its two energy eigenstates:

$$\psi_n = a_1(L)\phi_1(x) + a_2(L)\phi_2(x) \qquad 4$$

where $\phi_1$ and $\phi_2$ are the wavefunctions of the first and second states, respectively. The expectation value of the Hamiltonian is calculated using $E = \langle\psi|H|\psi\rangle$:

$$E(L) = \sum_{n=1}^{2}(|a_1|^2 + |a_2|^2)E_n = |a_1|^2 E_1 + |a_2|^2 E_2 \qquad 5$$

where the coefficients are constrained by the normalization condition $|a_1|^2 + |a_2|^2 = 1$ and $E_n$ ($n = 1,2,3,...$) maybe obtained from the [WS] energy spectrum. At $n = 1$ we have that the heat input $E_H$ into the system is

$$E_H = -\left(\frac{V_0}{2} + \frac{\hbar^2}{2mL_1^2} + \frac{V_0^2 mL_1^2}{8\hbar^2}\right) \qquad 6$$

Considering the transition from $n = 1$ to $n = 2$

$$E(L) = \left(\frac{5\hbar^2}{8mL^2} - \frac{5V_o^2 mL^2}{72\hbar^2}\right)|a_1|^2 - \left(\frac{V_0}{2} + \frac{9\hbar^2}{8mL^2} + \frac{V_o^2 mL^2}{18\hbar^2}\right) \qquad 7$$

Since the expectation value is constant during the process, we set the expectation value to be equal to $E_H$

$$-\left(\frac{V_0}{2} + \frac{\hbar^2}{2mL_1^2} + \frac{V_o^2 mL_1^2}{8\hbar^2}\right) = -\frac{V_0}{2} + \frac{\hbar^2}{8mL^2}[5|a_1|^2 - 9] - \frac{V_o^2 mL^2}{72\hbar^2}[5|a_1|^2 + 4] \qquad 8$$

Comparing coefficients and solving for $L$ we obtained:

$$L^2 = -\frac{L^2_1}{4}[5|a_1|^2 - 9]$$

$$L^2 = \frac{9L^2_1}{[5|a_1|^2 + 4]} \qquad 9$$

The maximum value of $L$ is obtained when $L = L_2$. This is at the end of the isothermal expansion process where $|a_1|^2 = 0$. Therefore, from eq. (9)

$$L_2^2 = \frac{9L_1^2}{4} \Rightarrow L_2 = \frac{3}{2}L_1 \qquad 10$$

The pressure during the isothermal expansion is:

$$P(L) = \sum_{n=1}^{\infty}|a_n|^2 P_n = |a_1|^2 P_1 + |a_2|^2 P_2 \qquad 11$$

The pressure undergoes a change of state from $n = 1$ to $n = 2$ as such resulting in:

$$P(L) = \frac{V_o^2 mL}{36\hbar^2}[5|a_1|^2 + 4] + \frac{\hbar^2}{4mL^3}[5|a_1|^2 - 9] \qquad 12$$

$$P(L) = \frac{V_o^2 mL}{9\hbar^2} - \frac{9\hbar^2}{4mL^3} \qquad 13$$

Substituting the value of (10) into (12) gives

$$P_1(L) = \frac{V_o^2 mL}{9\hbar^2} - \frac{\hbar^2}{mL_1^2 \cdot L} \qquad 14$$

The product $LP_1(L) = constant$. This is an exact quantum analogue of a classical *equation of state*.

### b. Process 2: Adiabatic Expansion

Next, the system then undergoes an adiabatic expansion from $L = L_2$ until $L = L_3$. During the process, the system maintains the second state at $n = 2$ as no external energy comes into the system. The expectation value of the Hamiltonian is:

$$E(L) = -\left(\frac{V_0}{2} + \frac{9\hbar^2}{8mL_2^2} + \frac{V_o^2 mL_2^2}{18\hbar^2}\right) = E_C \qquad 15$$

The pressure $P$ as a function of $L$ is:

$$P_2 = \frac{V_o^2 mL}{9\hbar^2} - \frac{9\hbar^2}{4mL^3} \qquad 16$$

The product $L^3 P_2(L)$ in (16) is a constant and it is considered the quantum analogue of the classical *adiabatic process*.

c. **Process 3: Isothermal Compression**

This process appears to be a reversal of process 1, the system is compressed isothermally from its current (second) state $n = 2$ at point 3 back to its initial state $n = 1$ in point 4 (i.e., from $L = L_3$ until $L = L_4$) while the expectation value of the Hamiltonian remains constant. Thus, the state of the system is a linear combination of its two energy eigenstates.

$$\Psi_n = b_1(L)\phi_1(x) + b_2(L)\phi_2(x)$$

where $\phi_1$ and $\phi_2$ are the wave functions of the first and second states respectively

$$E(L) = \sum_{n=1}^{\infty}(|b_1|^2 + |b_2|^2)E_n = |b_1|^2 E_1 + |b_2|^2 E_2 \qquad 17$$

The coefficients are constrained by the normalization condition $|b_1|^2 + |b_2|^2 = 1$. The expectation value of the Hamiltonian in this state as a function of $L$ is calculated using $E = \langle\psi|H|\psi\rangle$, which results in:

$$E(L) = -\frac{V_0}{2} - \frac{\hbar^2}{8mL^2}[4 - 5|b_2|^2] + \frac{V_o^2 mL^2}{72\hbar^2}[5|b_2|^2 - 9] \qquad 18$$

Setting the expectation value to be equal to $E_C$

$$E_C = -\left(\frac{V_0}{2} + \frac{9\hbar^2}{8mL_3^2} + \frac{V_o^2 mL_3^2}{18\hbar^2}\right) \qquad 19$$

Since the expectation value is constant during the process, set the expectation value is set to be equal to $E_L$ i.e., $n = 2$

$$-\left(\frac{V_0}{2} + \frac{9\hbar^2}{8mL_3^2} + \frac{V_o^2 mL_3^2}{18\hbar^2}\right) = -\frac{V_0}{2} - \frac{\hbar^2}{8mL^2}[4 - 5|b_2|^2] + \frac{V_o^2 mL^2}{72\hbar^2}[5|b_2|^2 - 9] \qquad 20$$

Comparing coefficients and solving for $L$ we obtain:

$$L^2 = -\frac{L_3^2}{9}[4 - 5|b_2|^2]$$

$$L^2 = \frac{4L_3^2}{[5|b_2|^2 - 9]} \qquad 21$$

The max value of $L$ is obtained when $L = L_4$ and this is achieved in the isothermal compression when $|b_2|^2 = 0$. Therefore, from eq. (21)

$$L_4^2 = \frac{4L_3^2}{9} \equiv L_4 = \frac{2}{3}L_3 \qquad 22$$

The pressure during the isothermal compression is:

$$P_3(L) = \frac{V_o^2 mL}{4\hbar^2} - \frac{9\hbar^2}{4mL_3^2 \cdot L} \qquad 23$$

The product $LP_3(L) = constant$. This is an exact quantum analogue of a classical *equation of state*

### d. Process 4: Adiabatic Compression

The system remains in its initial state $n = 1$ at point 4 as it undergoes an adiabatic compression process (i.e., from $L = L_4$ until $L = L_1$). The expectation value of the Hamiltonian is then given by:

$$E(L) = -\left(\frac{V_0}{2} + \frac{\hbar^2}{2mL^2} + \frac{V_o^2 mL^2}{8\hbar^2}\right) \qquad 24$$

and the pressure $P$ as a function of $L$ during this process is given as:

$$P_4(L) = \frac{V_o^2 mL}{4\hbar^2} - \frac{\hbar^2}{mL^3} \qquad 25$$

The product $L^3 P_4(L)$ in (25) is a constant and it is considered the quantum analogue of the classical *adiabatic process*.

### e. Work done in one closed cycle

During one closed cycle, the new work done $W$ by the quantum heat engine is described by the area of the closed loops gotten from the processes i.e., a summation of the individual work done by each process as shown in eqs (14), (16), (23) and (25).

$$W = W_{12} + W_{23} + W_{34} + W_{41} \qquad 26$$

$$W = \int_{L_1}^{L_2} P_1(L).dL + \int_{L_2}^{L_3} P_2(L).dL + \int_{L_3}^{L_4} P_3(L).dL + \int_{L_4}^{L_1} P_4(L).dL \qquad 27$$

Recall that: $L_2 = \left(\frac{3}{2}\right)L_1$ and $L_4 = \left(\frac{2}{3}\right)L_3$, therefore:

$$W = \int_{L_1}^{\frac{3}{2}L_1}\left(\frac{V_o^2 mL}{9\hbar^2} - \frac{\hbar^2}{mL_1^2.L}\right).dL + \int_{\frac{3}{2}L_1}^{L_3}\left(\frac{V_o^2 mL}{9\hbar^2} - \frac{9\hbar^2}{4mL^3}\right).dL$$
$$+ \int_{L_3}^{\frac{2}{3}L_3}\left(\frac{V_o^2 mL}{4\hbar^2} - \frac{9\hbar^2}{4mL_3^2.L}\right).dL + \int_{\frac{2}{3}L_3}^{L_1}\left(\frac{V_o^2 mL}{4\hbar^2} - \frac{\hbar^2}{mL^3}\right).dL \qquad 28$$

$$W = \left(\frac{V_o^2 mL_1^2}{72\hbar^2} - \frac{\hbar^2 \ln\left(\frac{3}{2}\right)}{mL_1^2}\right) - \left(\frac{V_o^2 mL_3^2}{72\hbar^2} - \frac{9\hbar^2 \ln\left(\frac{3}{2}\right)}{4mL_3^2}\right) \qquad 29$$

By definition, the efficiency $\eta$ of a heat engine is defined as.

$$\eta = \frac{W}{Q_H} \qquad 30$$

given that $Q_H$ is the quantity of heat in the hot reservoir and $W$ is the work performed by the classical heat engine which is analogous to the energy absorbed by the quantum engine during the isothermal expansion:

$$Q_H = \int_{L_1}^{L_2} P_1(L).dL \qquad 31$$

$$Q_H = \left(\frac{V_o^2 mL_1^2}{72\hbar^2} - \frac{\hbar^2 \ln\left(\frac{3}{2}\right)}{mL_1^2}\right) \qquad 32$$

Therefore, the efficiency $\eta$ of a quantum heat engine obtained by considering a two-state system in a WS potential model is given by;

$$\eta = 1 - \frac{\left(\frac{V_o^2 mL_3^2}{72\hbar^2} - \frac{9\hbar^2 \ln\left(\frac{3}{2}\right)}{4mL_3^2}\right)}{\left(\frac{V_o^2 mL_1^2}{72\hbar^2} - \frac{\hbar^2 \ln\left(\frac{3}{2}\right)}{mL_1^2}\right)} \qquad 33$$

which may be rewritten as

$$\eta = 1 - \left(\frac{9L_1^2}{4L_3^2}\right)\left(\frac{\alpha - 1}{\beta - 1}\right) \qquad 34$$

where

$$\alpha = \frac{4V_o{}^2 m^2 L_3^4}{9\left(72\hbar^4 \ln\left(\frac{3}{2}\right)\right)} \quad \text{and} \quad \beta = \frac{V_o{}^2 m^2 L_1^4}{72\hbar^4 \ln\left(\frac{3}{2}\right)}$$

Eq. (34) shows that the efficiency obtained with the more realistic WS model takes the same form as in the case of the harmonic oscillator and the free particle models [24,29,30] albeit with a richer insight depending on the magnitude of the factor $9(\alpha - 1)/4(\beta - 1)$.

## 3. RESULTS AND DISCUSSION

It is well known that the [WS] potential has been widely used in describing nuclear processes. The recent application of the interaction to low-dimension systems has equally been successful [40–44,48], hence the motivation for the present application to [QHE]. An advantage of the more realistic [WS] interaction is the interaction width $L$ which mimics the wall of the classical heat engine is well defined and explicitly parameterized unlike the case of the well-studied oscillator interaction. To validate the derived Carnot's efficiency of the Woods-Saxon [WS] model in Eq. (33), the present result is compared with earlier works which similarly modelled the Carnot's engine using the free particle and Pöschl-Teller interactions within appropriate limits [24,30]. From Eq. (34), depending on the relative values of $L_1^2$ and $L_3^2$, it is observed that the factor $(\alpha - 1)/(\beta - 1)$ may be greater or lesser than one (i.e., $\alpha > \beta$ or $\alpha < \beta$ respectively). Thus, giving the lower and upper limits of the possible efficiency. However, considering the order of magnitude of $\alpha$ and $\beta$ i.e. ($\alpha \gg 1$ and $\beta \gg 1$), it is safe to assume that the factor $(\alpha - 1)/(\beta - 1) \sim 1$, hence

$$\eta = 1 - \frac{9L_1^2}{4L_3^2} \qquad 35$$

We note also that Eq. (35) agrees with the result for free particle model in which the depth $V_o = 0$.

Substituting Eqs. (6) and (19) into (35), generates an efficiency

$$\eta = 1 - \frac{E_C}{E_H} \qquad 36$$

which is analogous to the classical Carnot's efficiency since $E_C \cong T_C$ and $E_H \cong T_H$. [24,30].

## 4. CONCLUSION

In this work, we have shown explicitly that the [WS] potential can successfully be used in modeling the nonlinearities of a [QHE]. We derived the efficiency of a cyclic engine using the Eigen energies of a non-linear quantum system described by the Wood-Saxon [WS] potential model. The cycles are shown to be analogous to their respective classical isothermal and adiabatic processes of the Carnot cycle and the efficiency obtained with the [WS] model is found to agree with existing results within appropriate limits. Extension of the present model to more realistic cycles and the finite-time processes, to be published elsewhere, have shown further the advantages of the realistic interaction in describing the promising results.